\documentclass[a4paper,fleqn,usenatbib]{mnras}

\usepackage{mathptmx}
\usepackage[T1]{fontenc}
\usepackage{ae,aecompl}
\usepackage{placeins}  % For \FloatBarrier

\usepackage{graphicx}	% Including figure files
\usepackage{amsmath}	% Advanced maths commands
\usepackage{amssymb}	% Extra maths symbols

\usepackage{multirow}
\usepackage[flushleft]{threeparttable}

\title[H/D–CO$_2$ superthermal collisions]
{Quantum scattering of hot H/D on CO$_2$: Cross sections and rate coefficients for planetary atmospheres and their evolution}

\author[Bop \& Gacesa]{
Cheikh T. Bop$^{1}$ \&
Marko Gacesa$^{1,2}$\thanks{E-mail: marko.gacesa@ku.ac.ae}
\\
$^{1}$
Physics Department, Khalifa University, Abu Dhabi, United Arab Emirates \\
$^{2}$ 
Khalifa University Space Technology \& Innovation Laboratory, Khalifa University, Abu Dhabi, United Arab Emirates
}

\date{Uploaded on \today}

\pubyear{2025}

\begin{document}
\label{firstpage}
\pagerange{\pageref{firstpage}--\pageref{lastpage}}
\maketitle

\begin{abstract}
Collisions between hot hydrogen atoms and CO$_2$ play a central role in energy transfer and atmospheric
escape in CO$_2$-rich planetary atmospheres. We present quantum mechanical $j_z$-conserving coupled-states
calculations of state-resolved cross sections for H/D--CO$_2$ collisions at energies up to 5~eV, benchmarked
to within 7\% of close-coupling results. Scattering is strongly forward-peaked, yielding momentum-transfer
cross sections substantially smaller than commonly assumed: mass-scaling from O/C--CO$_2$ systems
overestimates H--CO$_2$ total cross sections by factors of 30--45, while existing empirical fits
underestimate the low-energy regime by up to $\sim$45\%. Isotopic substitution (H/D) produces
energy-dependent differences of up to 35\% at $E<0.1$~eV, invalidating uniform scaling approaches for
D/H fractionation. Maxwellian-averaged rate coefficients derived from our cross sections are significantly
smaller than mass-scaled values, implying reduced H--CO$_2$ energy transfer efficiency. In atmospheric
escape modelling, these revisions can shift Martian exobase altitudes by 10--20~km, leading to order-unity
changes in thermal escape rates, and have implications for hydrogen loss in early CO$_2$-dominated
planetary atmospheres. Our results provide essential quantum-mechanical inputs for revisiting atmospheric
evolution scenarios on Mars, early Earth, and CO$_2$-rich exoplanets.
\end{abstract}

% Select between one and six entries from the list of approved keywords.
% Don't make up new ones.

%%%%%%%%%%%%%%%%% BODY OF PAPER %%%%%%%%%%%%%%%%%%

\section{Introduction}

Carbon dioxide (CO$_2$) is among the most abundant molecules in planetary, cometary, and exoplanetary atmospheres \citep{feaga2007asymmetries,LimayeVenus2018,Madhusudhan_2012,turbet2018modeling,Cadieux_2024}, including temperate terrestrial worlds like TRAPPIST-1e and LHS 1140 b where it may dominate atmospheric composition.
Its large moment of inertia and high atmospheric abundance make CO$_2$ a dominant collision partner in the upper atmospheres of terrestrial planets, where it governs radiative cooling and mediates energy exchange with suprathermal atoms. 

Modelling hydrogen escape from CO$_2$-rich atmospheres has historically relied on simplified 
collisional treatments dating back to foundational work by \citet{hunten1973escape} and 
\citet{schunk1980ionospheres}, which employ hard-sphere or isotropic scattering approximations. 
These approaches underpinned the development of hydrodynamic escape theory and its application to 
mass fractionation of noble gases during the 'flight of the nobility' and water loss from 
terrestrial planets \citep{ZAHNLE1986462,ZAHNLE1990502}. 
However, these classical approximations do not capture the quantum-mechanical details of 
light-atom--molecule collisions --- particularly the strongly forward-peaked scattering that 
characterizes H/D--CO$_2$ interactions.
In the absence of system-specific data, modern studies often resort to reduced-mass scaling or surrogate systems \citep{lewkow2014precipitation,gacesa20203}, despite evidence that such approximations can introduce order-of-magnitude errors in transport cross sections. Moreover, for light atoms colliding with heavy polyatomic targets, quantum diffraction and interference dominate the angular distribution, rendering classical or isotropic approximations particularly unreliable.

On Mars, where CO$_2$ constitutes over 95\% of the present atmosphere \citep{jakosky2017loss}, these uncertainties propagate directly into atmospheric escape and isotopic fractionation calculations. Photochemical production of energetic H atoms drives long-term water loss and surface oxidation \citep{jakosky2018loss,Gregory2023Nonthermal}, creating a substantial suprathermal population capable of escaping from the upper thermosphere \citep{Amerstorfer201escape,fox2009photochemical,Jane2015chemistry}. These atoms arise primarily from dissociative recombination of ions such as HCO$^+$ \citep{fox2009photochemical,Gregory2023hco+}.
Observations reveal strong seasonal and solar-cycle modulation of hydrogen escape \citep{Chaffin2014,Mayyasi2023,susarla2024variability}, linked to perihelion heating, water transport, and exobase H densities. Crucially, MAVEN data show that escape of both H and D is controlled not only by thermal processes but also by a persistent suprathermal component \citep{Clarke2024,Lillis2017}. A purely thermal escape rate for D is nearly two orders of magnitude too small to explain the observed upper-atmospheric D/H variations. In this regime, isotopic fractionation becomes highly sensitive to the efficiency of momentum and energy loss through collisions with CO$_2$, making accurate elastic and rotationally inelastic H/D--CO$_2$ cross sections essential for interpreting modern D/H ratios in terms of long-term water loss.
Recent photochemical models confirm that deuterium escape requires a significant non-thermal component \citep{Cangi2023}, and isotopic evolution studies show that preferential escape of lighter isotopes controls the modern isotopic composition of the Martian atmosphere \citep{Thomas2023}. In the upper thermosphere, suprathermal H and D atoms experience only a handful of collisions, so the assumed elastic and momentum-transfer cross sections directly govern escape probabilities and D/H fractionation \citep{Gregory2023hco+,chaufray2024simulations}.

Here we present new quantum mechanical calculations of state-to-state, total, and differential cross sections for rotationally elastic and inelastic collisions of H and D atoms with CO$_2$ at collision energies up to 5~eV. These results provide essential inputs for modelling suprathermal energy transfer, hydrogen and deuterium escape, and isotopic evolution in CO$_2$-rich planetary atmospheres. Section~2 describes the computational approach, Section~3 presents the results and discussion, and Section~4 summarizes the main conclusions.

\section{Computational details}
\label{sec:comp}
\subsection{Potential energy surface}
\label{sec:pes}

The interaction between H and rigid-rotor CO$_2$ is described by the high-level \textit{ab initio} potential energy surface (PES) of \cite{dagdigian2015accurate}.
The PES was computed using the coupled-cluster method RCCSD(T) with an augmented quadruple-zeta basis set and includes corrections for basis-set superposition error.
We employ the same PES for both H--CO$_2$ and D--CO$_2$ systems, as the Born--Oppenheimer interaction depends only on nuclear coordinates and is invariant under isotopic substitution.
For details of the PES construction and its contour plots, we refer the reader to the original paper \citep{dagdigian2015accurate}.

\subsection{Cross sections}
We implemented the radial coefficients ($V_{\lambda}$) of the H-CO$_2$ PES, made available by \cite{dagdigian2015accurate}, in the \texttt{MOLSCAT} computer code \citep{hutson1994molscat}. The long-range part ($R\geq 28~a_0$) of these coefficients was derived by extrapolation using the following inverse power law \citep{ndaw2021excitation,bop2023collisional}:
\begin{eqnarray}
    V_{\lambda}(R) = \frac{C_{\lambda}}{R^{\eta_{\lambda}}}.
\end{eqnarray}
The parameters $C_{\lambda}$ and $\eta_{\lambda}$ were retrieved using the radial coefficients corresponding to the last {\it ab initio} points to ensure a gradual descent of the potentials. 
The scattering processes of interest in this study can be described as
\begin{eqnarray}
{\rm CO_2}(j) + {\rm H/D} & \to & {\rm CO_2}({j'}) + {\rm H/D},
\end{eqnarray}
where $j$ denotes the rotational quantum number of CO$_2$. Since the projectiles have kinetic energies reaching up to 5~eV, we employed the quantum mechanical $j_z$-conserving coupled states (CS) approximation \citep{mcguire1974quantum}, combined with the hybrid log-derivative-Airy propagator of \cite{alexander1987stable}, to solve the scattering problem. The CS method offers significant computational savings compared to the fully rigorous close-coupling (CC) approach \citep{alexander1977close} and remains accurate at suprathermal collision energies, typically above $\sim$0.1--0.2~eV  \citep{gong2025shock}.

The scattering calculations were performed for total energies up to $E=5$~eV. To derive highly accurate collision data, we performed prior convergence tests. The rotational basis was set large enough, including the 141 low-lying energy levels for CO$_2$, i.e. $j = 0-140$. The integration boundaries, $R_{\rm min}$ and $R_{\rm max}$, were automatically adjusted for each total angular momentum ($J$) of the collision systems H-CO$_2$ and D-CO$_2$. We used up to 241 partial waves, i.e. $J=0-240$, to achieve convergence of both elastic and inelastic cross sections. The parameter \texttt{STEPS}, which is interpreted as the number of steps per half-wavelength for the open channel of highest kinetic energy in the asymptotic region, was set to 30 for $E\leq1$~eV and 10 for total energies up to 5~eV. All these parameters ensure the subpercent convergence of high-magnitude state-to-state cross sections ($\geq10^{-2}$ $\mathring{A}^2$). The low-magnitude collision data have very minor effect on the total cross sections, and their subpercent convergence requires an extremely large rotational basis ($j = 0-350$), which is computationally challenging due to memory limitations and CPU times.
\paragraph*{Scope and limitations.}
%\subsection{Scope and limitations}
The present calculations treat CO$_2$ as a linear rigid rotor with fixed bond lengths, so vibrational excitation and vibration–rotation coupling are neglected. This approximation is suitable for the cross sections presented here because, over the energy range considered, vibrational excitation mainly redistributes energy into internal degrees of freedom of CO$_2$ and is expected to have a sub-dominant effect on the momentum-transfer cross sections most relevant to atmospheric escape.

The interaction is described by a single ground-state Born--Oppenheimer potential energy surface, so electronic
excitation and non-adiabatic effects are not included. Finally, most scattering calculations employ the $j_z$-conserving
coupled-states approximation rather than full close coupling; based on the benchmark in Fig.~\ref{fig_TXS}, the
resulting transport cross sections are accurate to within $\sim$7\% over the energy range considered, but the CS
approximation may be less reliable very near threshold and for subtle interference features in differential cross sections.
We also neglect reactive channels and treat the collisions as non-reactive scattering on a single PES; the results are intended for energy and momentum transfer in atmospheric-transport contexts rather than chemical kinetics, and earlier studies of reactive scattering involving O($^{3}P$) and H$_2$ at suprathermal energies reported reactive cross sections at least two orders of magnitude smaller than for non-reactive channels \citep{balakrishnan2004quantum,2003JChPh.118.1585G,2014JChPh.141p4324G,Gacesa_2017-OH}.

\section{Results and discussion}
\label{sec:results}

\subsection{Integral and differential cross sections}

\begin{figure}
    \centering
    \includegraphics[width=0.98\linewidth, trim = 4 0 45 30, clip = true]{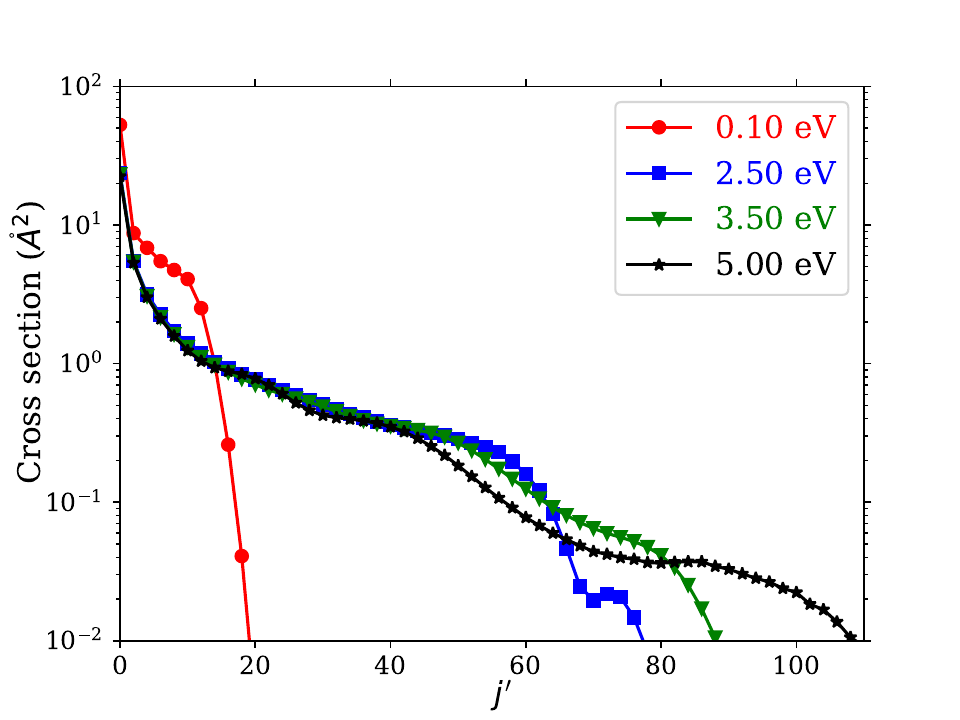}
    \caption{State-to-state cross sections $\sigma_{j=0 \to j'}$ for the scattering of CO$_2$ with H for selected collision energies.}
    \label{fig_state_to_state_XS}
\end{figure}

Fig.~\ref{fig_state_to_state_XS} shows the state-to-state cross sections $\sigma_{j=0\to j'}$ for selected collision energies, representative of translationally hot H atoms colliding with thermal CO$_2$. Throughout this section, collision energies are expressed in the centre-of-mass frame unless stated otherwise.
The contribution of inelastic transitions decreases rapidly with increasing $j'$. At all energies considered, the largest $\Delta j$ transitions are up to four orders of magnitude smaller than the elastic channel and up to three orders of magnitude smaller than the dominant inelastic transitions. Transitions to $j' > 140$, which are not included in this work, therefore contribute negligibly to the total cross section $\sigma_{j=0}^{\rm tot}$.
\begin{figure}
    \centering
    \includegraphics[width=0.98\linewidth, trim = 18 22 53 25, clip = true]{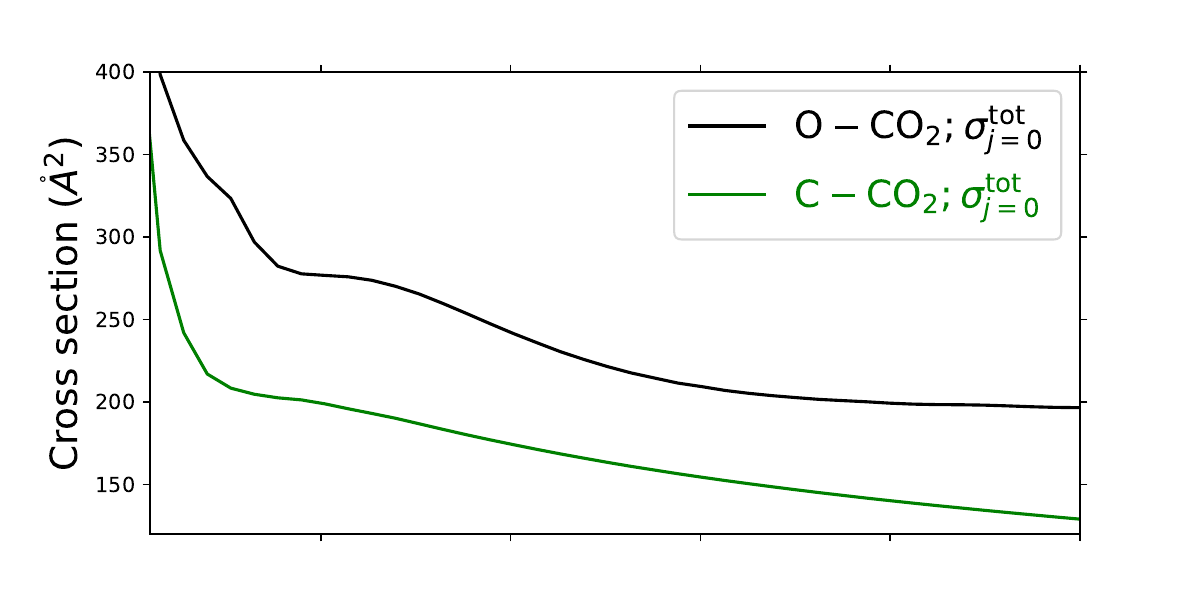}    
    \includegraphics[width=0.98\linewidth, trim = 16 22 55 55, clip = true]{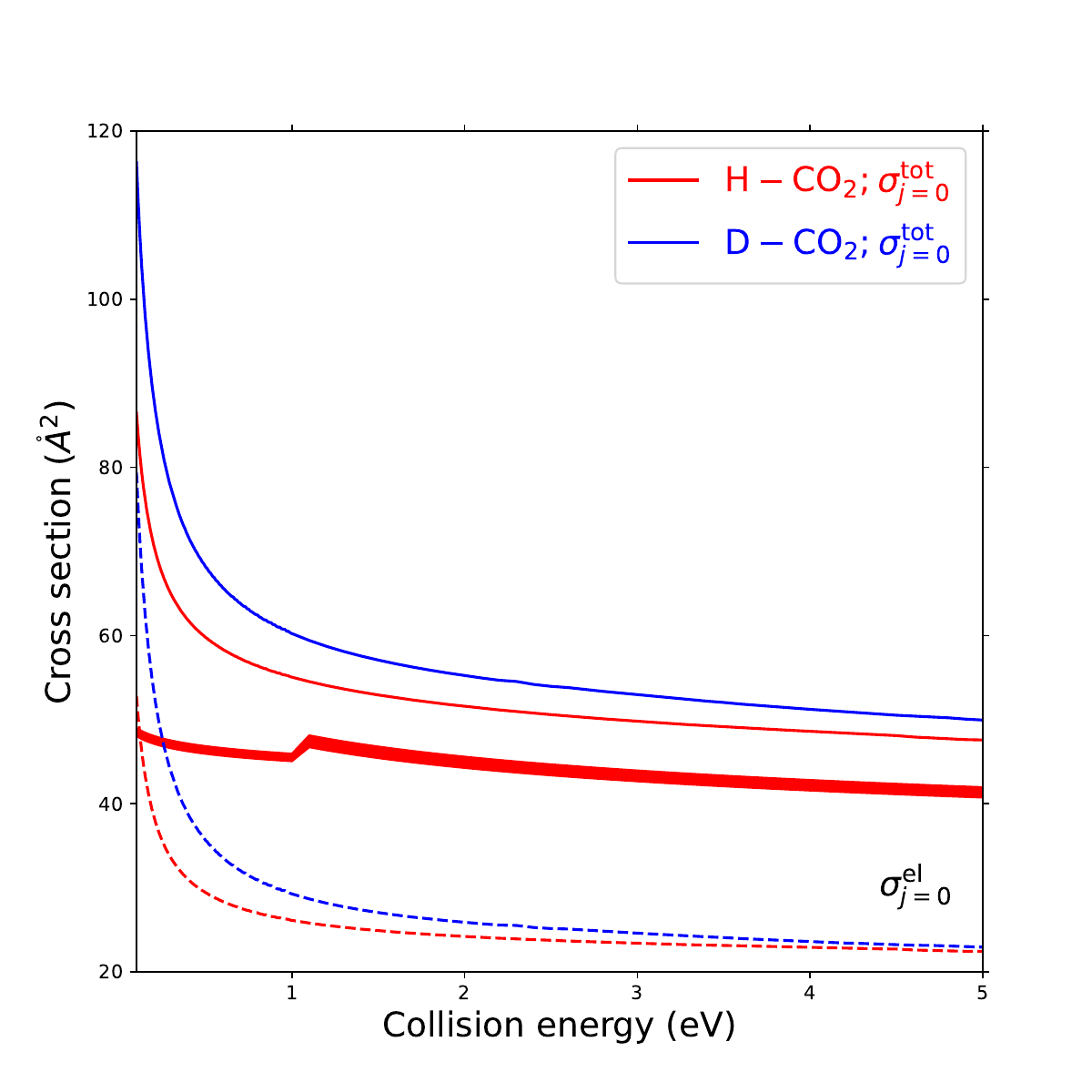}
    \caption{Total cross sections $\sigma_{j=0}^{\rm tot}$ for CO$_2$ in collisions with H, D, O, and C as a function of collision energy. The O--CO$_2$ and C--CO$_2$ results are shown after reduced-mass scaling for comparison.
The thick red line denotes the error margin of the H--CO$_2$ total cross sections from \citet{lewkow2014precipitation}.
Dashed red and blue lines show the elastic contributions for H--CO$_2$ and D--CO$_2$, respectively.}
    \label{fig_XS_tot}
\end{figure}

\begin{figure*}
    \centering
    \includegraphics[width=0.48\linewidth, trim = 13 13 13 13, clip = true]{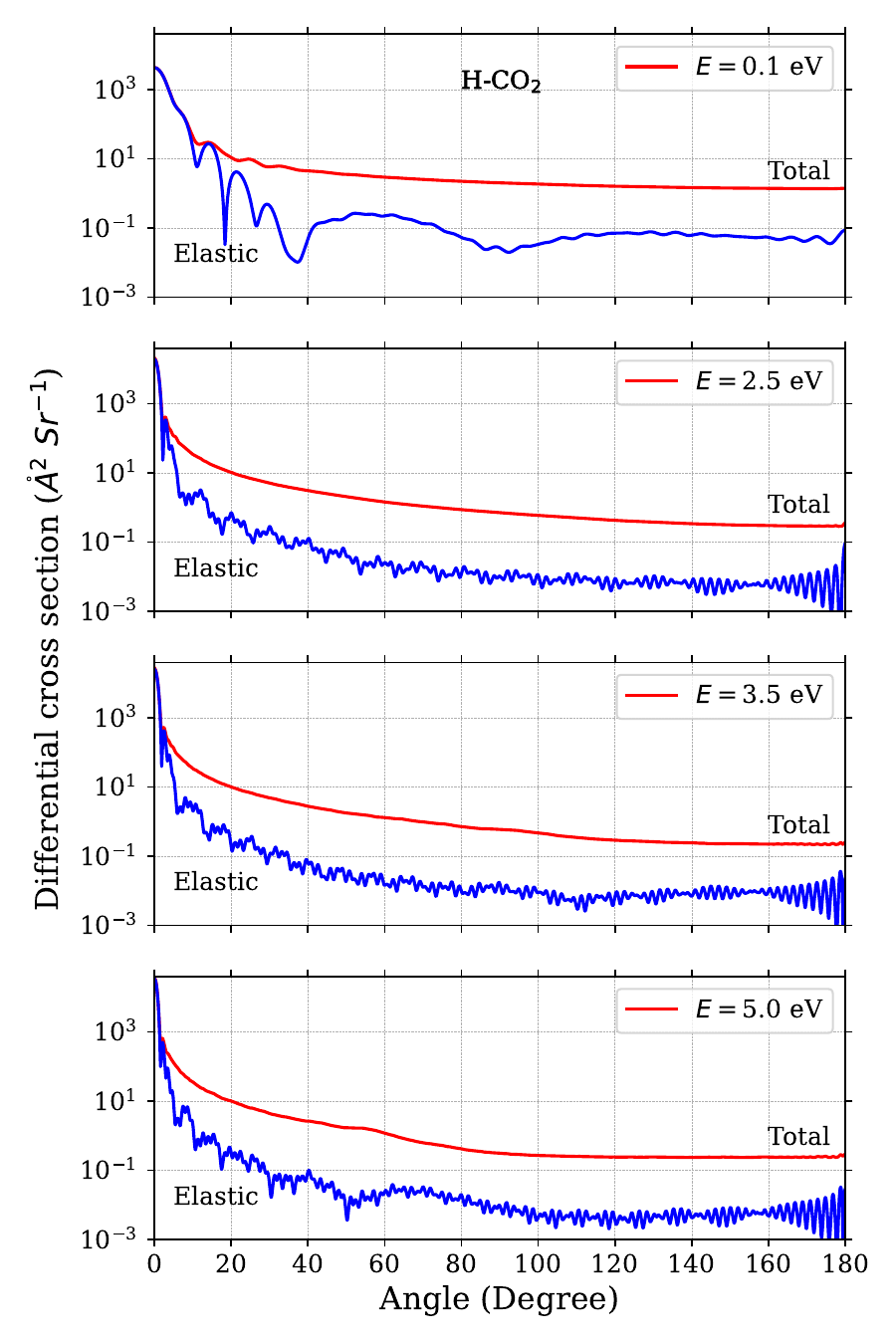}
    \includegraphics[width=0.48\linewidth, trim = 13 13 13 13, clip = true]{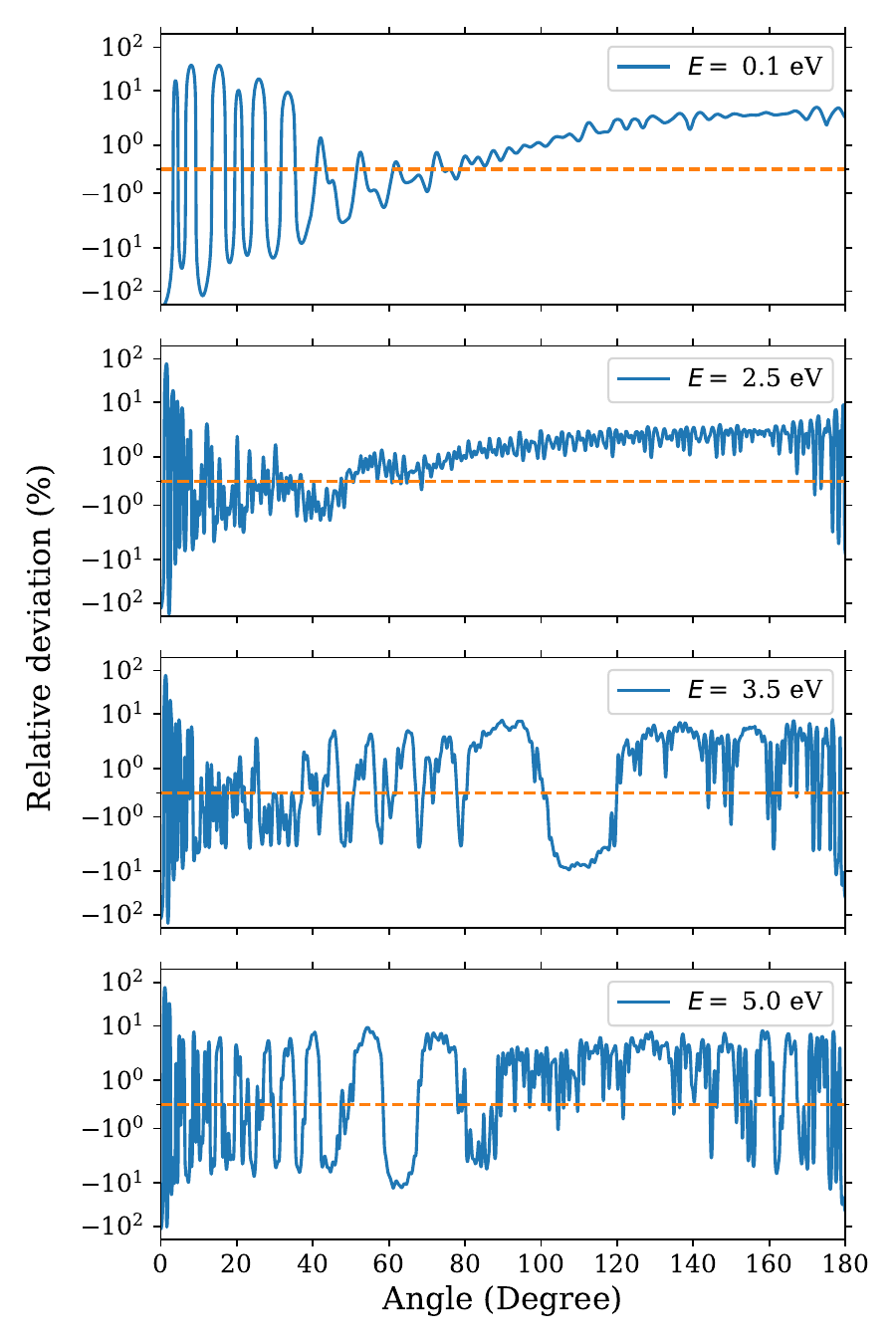}
    \caption{Angular dependence of H--CO$_2$ differential cross sections (left panel) and relative deviations upon H/D isotopic substitution (right panel) for selected collision energies. The horizontal line marks perfect agreement. Large relative deviations at very small angles occur in a strongly forward-peaked regime and should be interpreted together with the corresponding absolute cross sections.}
    \label{fig_DXS}
\end{figure*}

In Fig.~\ref{fig_XS_tot} we compare total cross sections $\sigma_{j=0}^{\rm tot}$ for CO$_2$ in collisions with different projectiles for kinetic energies up to 5~eV. As seen in the lower panel, the H--CO$_2$ and D--CO$_2$ cross sections follow similar energy-dependent trends, with differences decreasing as the kinetic energy increases. At $E \leq 0.1$~eV, D-induced cross sections exceed those from H by up to $\sim 35$\%, whereas for $E > 0.1$~eV the difference falls below 9\%. This reflects the longer de Broglie wavelength and greater quantum diffractive scattering for the lighter H projectile at low energies.
For both isotopes, elastic scattering accounts for 40-55\% of the total cross section across the full energy range.

The upper panel of Fig.~\ref{fig_XS_tot} compares H--CO$_2$ total cross sections with those for O--CO$_2$ and C--CO$_2$ computed previously \citep{gacesa20203,gacesa2024elastic}. Even after reduced-mass scaling, these heavier projectiles produce cross sections that are substantially larger than the true H--CO$_2$ values. Using O--CO$_2$ (C--CO$_2$) as a proxy leads to typical scaling factors of $\sim 12$ ($\sim 9.5$) and overestimation of the actual H--CO$_2$ total cross section by up to factors of $\sim 45$ ($\sim 30$). Such discrepancies highlight the limitations of mass-scaling approaches frequently employed in atmospheric escape modelling.
The scaled differential cross section fit of \citet{lewkow2014precipitation} reproduces the general behaviour of the H--CO$_2$ total cross sections above $\sim 1$~eV, with deviations at the 10-15\% level. At lower energies, however, where quantum effects are more pronounced, the deviation increases to $\sim 45$\%. Although the agreement at high energies is encouraging, the residual discrepancies remain larger than the isotopic H/D differences, reinforcing the need for dedicated quantum calculations for light-atom projectiles. This is expected because the fitting approach does not explicitly resolve the full state-to-state rotational dynamics of the H--CO$_2$ system.

The left panel of Fig.~\ref{fig_DXS} shows the angular dependence of the total and elastic differential cross sections for H--CO$_2$. Both exhibit strong forward scattering, with the forward peak contributing more than 90\% of the integral cross section. This behaviour intensifies at higher collision energies, while the differential cross section decreases sharply with energy for scattering angles above $\sim 20^\circ$. The elastic contribution falls rapidly with angle: it exceeds 80\% below 2$^\circ$ but drops below $\sim 5$\% at 180$^\circ$ across all energies considered.

The right panel of Fig.~\ref{fig_DXS} shows the relative deviations in differential cross sections between H and D projectiles. As observed previously \citep{Bop2025energy,10.1093/mnras/stac3057,zhang2009energy}, the deviations are highly anisotropic and largest at small angles, where the differential cross section is most forward-peaked. For scattering angles below $\sim2^\circ$, fractional differences can reach two orders of magnitude at very small angles where the absolute cross section is already sharply peaked, whereas beyond this range the absolute deviations remain below $\sim50$\%. This complex angular structure demonstrates that isotopic substitution cannot be represented by a single mass-scaling correction.

\subsection{Transport parameters}

\begin{figure}
    \centering
    \includegraphics[width=0.98\linewidth, trim = 13 0 34 38, clip = true]{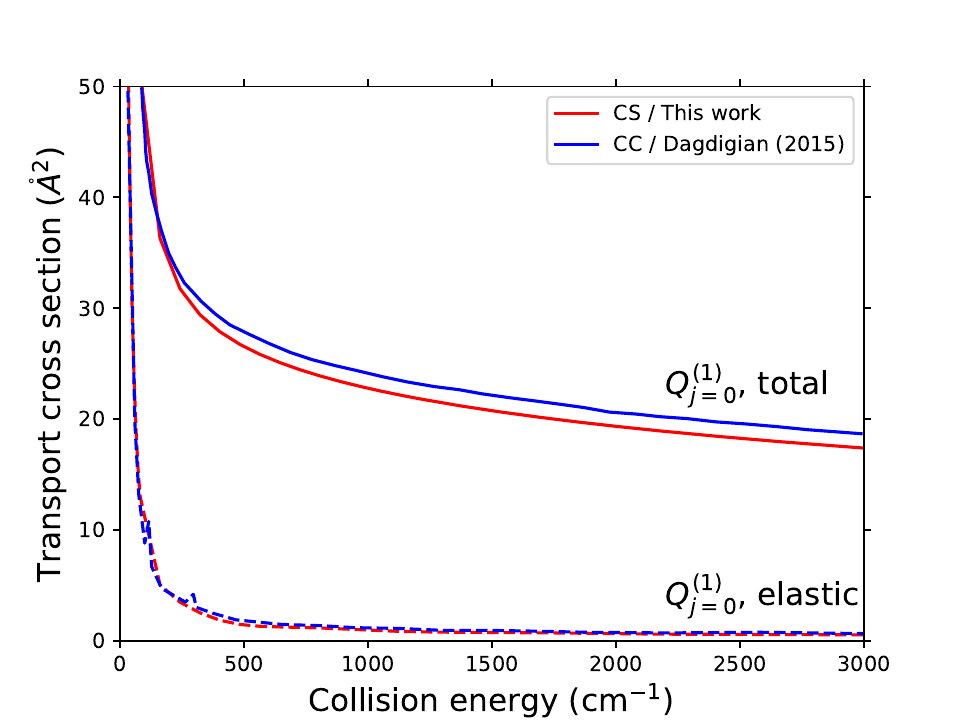}
    \caption{Dependence of state-specific transport cross sections on collision energy for H--CO$_2$($j=0$). Dashed lines show the elastic contribution. Comparison between close-coupling and coupled-states calculations.}
    \label{fig_TXS}
\end{figure}

Figure~\ref{fig_TXS} compares state-specific transport cross sections computed in this work using the CS approximation with the CC results of \citet{dagdigian2015accurate}. These cross sections weight collisions by $(1-\cos\theta)$ and quantify the efficiency of momentum exchange rather than the overall scattering probability \citep{schunk2000ionospheres}. For escape and thermalisation modelling, $\sigma_{\rm mt}$ is the relevant quantity because it controls momentum loss per collision. \footnote{The CC method, the most rigorous quantum approach, has itself been estimated to be accurate to within $\sim$10\% when compared with available laboratory data at higher collision energies \citep{wiesenfeld2025ab}.}
The two methods agree to within 7\% across the full energy range. These small deviations likely reflect differences in numerical convergence, as both calculations employ the same potential energy surface. 

For both isotopes, $\sigma^{\rm mt}$ is much smaller than the corresponding total and elastic cross sections across all energies, demonstrating that the scattering is strongly forward-peaked. As a result, individual H--CO$_2$ and D--CO$_2$ collisions rarely redirect the projectile by large angles, despite a high collision frequency. The small $\sigma_{\rm mt}/\sigma_{\rm tot}$ ratios imply long effective momentum-transfer mean free paths and inefficient collisional thermalisation, so hot H and D atoms retain a significant fraction of their incident momentum after each collision. As discussed in Sect.~4, this behaviour has direct implications for exobase structure and for the probability that suprathermal atoms escape or become quenched.

\subsection{Maxwellian-averaged rate coefficients}
\label{sec:ratecoeff}

For modelling applications, the relevant quantity is often the temperature-dependent rate coefficient rather than the energy-dependent cross section. We therefore compute Maxwellian-averaged momentum-transfer rate coefficients for thermal distributions of H and D atoms colliding with CO$_2$. The rate coefficient at temperature $T$ is obtained by integrating the state-specific momentum-transfer cross section $\sigma^{\rm mt}(E)$ over a Maxwell–Boltzmann energy distribution:

\begin{equation}
k_{\rm mt}(T) = \langle \sigma^{\rm mt} v \rangle_T = \sqrt{\frac{8}{\pi \mu (k_{\rm B}T)^3}} \int_{0}^{\infty} \sigma^{\rm mt}(E) E e^{-E/k_{\rm B}T} dE,
\label{eq:rate_coeff}
\end{equation}
where $\mu$ is the reduced mass of the collision system, $k_{\rm B}$ is Boltzmann's constant, $E$ is the collision energy in the centre-of-mass frame, and $v = \sqrt{2E/\mu}$ is the relative speed. The integration is performed numerically using Simpson's rule, with $\sigma^{\rm mt}_{j=0}(E)$ taken from the $j=0$ state-specific results shown in Fig.~\ref{fig_TXS}. The same potential energy surface is used for both isotopes, so any differences in $k_{\rm mt}(T)$ arise solely from the different reduced masses and quantum scattering dynamics.

\FloatBarrier % Force table to appear here

\begin{table}
\centering
\caption{Maxwellian-averaged momentum-transfer rate coefficients $k_{\rm mt}(T)$ for H/D--CO$_2$ collisions. The final column gives the isotopic ratio $k_{\rm D}(T)/k_{\rm H}(T)$.}
\label{tab:rate_coeff}
\begin{tabular}{cccc}
\hline
$T$ (K) & $k_{\rm H}$ (cm$^3$ s$^{-1}$) & $k_{\rm D}$ (cm$^3$ s$^{-1}$) & $k_{\rm D}/k_{\rm H}$ \\
\hline
100  & $3.53 \times 10^{-11}$ & $2.08 \times 10^{-11}$ & 0.589 \\
200  & $2.36 \times 10^{-11}$ & $1.46 \times 10^{-11}$ & 0.618 \\
300  & $1.84 \times 10^{-11}$ & $1.14 \times 10^{-11}$ & 0.619 \\
500  & $1.38 \times 10^{-11}$ & $8.41 \times 10^{-12}$ & 0.610 \\
1000 & $1.01 \times 10^{-11}$ & $5.97 \times 10^{-12}$ & 0.590 \\
2000 & $8.35 \times 10^{-12}$ & $4.81 \times 10^{-12}$ & 0.576 \\
3000 & $7.77 \times 10^{-12}$ & $4.44 \times 10^{-12}$ & 0.571 \\
5000 & $7.21 \times 10^{-12}$ & $4.09 \times 10^{-12}$ & 0.568 \\
\hline
\end{tabular}
\end{table}

Table~\ref{tab:rate_coeff} lists $k_{\rm mt}(T)$ for both isotopes over the temperature range 100--5000~K, which spans conditions relevant to the Martian thermosphere and early planetary atmospheres. Over this range, the rate coefficients increase by approximately a factor of 7, reflecting the energy dependence of $\sigma_{\rm mt}(E)$. 
The isotopic ratio $k_{\rm D}/k_{\rm H}$ is approximately 0.6 across the temperature range, 
indicating that deuterium atoms transfer about 40\% less momentum per collision than hydrogen 
atoms at comparable temperatures. This mass dependence is consistent with expectations for 
light-atom scattering where the reduced mass affects both the de Broglie wavelength and the 
classical momentum exchange.

\begin{figure}
\centering
\includegraphics[width=0.98\linewidth]{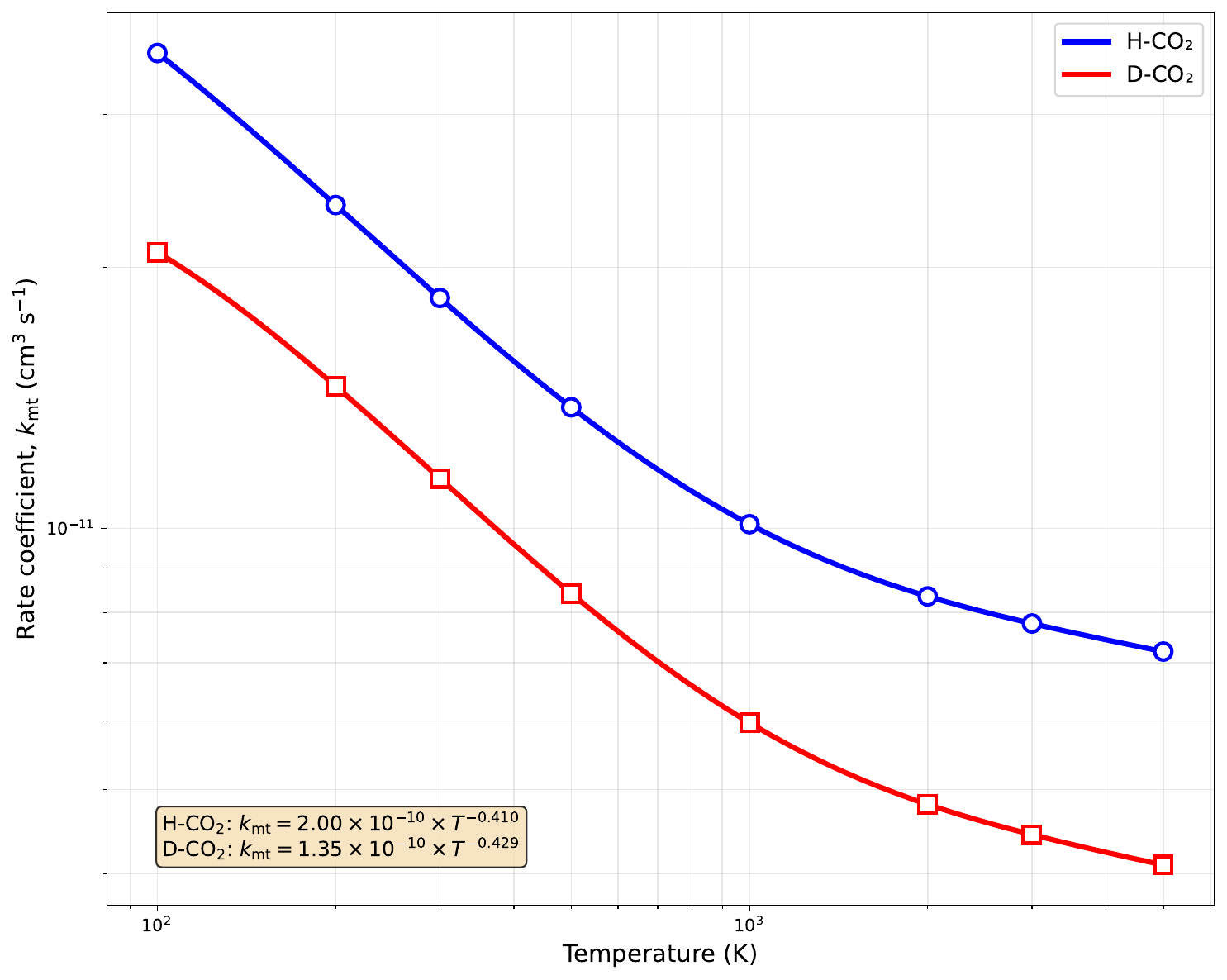}
\caption{
Maxwellian-averaged momentum-transfer rate coefficients $k_{\rm mt}(T)$ for H--CO$_2$ (blue circles and line) and D--CO$_2$ (red squares and line) collisions.
Solid curves show the smoothly interpolated values, while symbols mark the temperatures listed in Table~\ref{tab:rate_coeff}.
The behaviour is well approximated by power laws (dashed lines) of the form $k(T) = A T^{B}$ with 
$A_{\rm H}=2.00 \times 10^{-10}$~cm$^3$~s$^{-1}$, $B_{\rm H}=-0.410$ for H--CO$_2$ 
and $A_{\rm D}=1.35 \times 10^{-10}$~cm$^3$~s$^{-1}$, $B_{\rm D}=-0.429$ for D--CO$_2$.
}
\label{fig:rate_coeff}
\end{figure}

Figure~\ref{fig:rate_coeff} displays the temperature dependence of $k_{\rm mt}(T)$ for both isotopes. The curves are well described by simple power-law fits,
$k(T) = A T^{B}$, with the parameters given in the caption. These fits provide a convenient analytic representation for implementation in atmospheric models where a full integration of Eq.~(\ref{eq:rate_coeff}) is impractical.

The momentum-transfer rate coefficients presented here, $k_{\rm H}(300~{\rm K}) = 1.84 \times 10^{-11}$~cm$^3$~s$^{-1}$ and 
$k_{\rm D}(300~{\rm K}) = 1.14 \times 10^{-11}$~cm$^3$~s$^{-1}$, are consistent in magnitude with values for similar 
light-atom--molecule systems such as H--N$_2$ \citep{schunk2000ionospheres}. However, they differ significantly from 
values sometimes adopted in atmospheric models that employ simplified scaling approaches. For comparison, mass-scaling 
from O--CO$_2$ cross sections \citep{gacesa20203} would yield $k_{\rm H}(300~{\rm K}) \sim 3 \times 10^{-10}$~cm$^3$~s$^{-1}$, 
approximately 16 times larger than our quantum-mechanical result. This discrepancy arises because forward-peaked 
scattering characteristic of light-atom collisions yields much smaller momentum-transfer cross sections than 
would be predicted by isotropic or hard-sphere models.

\section{Conclusion}
\label{sec:conc}

We have computed comprehensive state-to-state, total, and transport cross sections for rotationally elastic
and inelastic collisions between hot hydrogen atoms and CO$_2$ using a time-independent,
$j_z$-conserving coupled-states quantum mechanical approach.
The calculations span collision energies relevant to both thermal and non-thermal escape processes in CO$_2$-rich planetary atmospheres, including the Martian upper atmosphere.

For CO$_2$ in its ground rotational state, inelastic cross sections decrease rapidly with increasing final
rotational quantum number $j'$, with transitions beyond $\Delta j \gtrsim 140$ contributing negligibly to
the total.
A systematic comparison of H--CO$_2$ and D--CO$_2$ collisions shows that isotopic differences are strongly
energy dependent: D--CO$_2$ cross sections exceed H--CO$_2$ values by up to $\sim$35\% at $E<0.1$~eV, while
differences fall below $\sim$9\% at higher energies.
These results demonstrate that isotopic substitution cannot be represented by a uniform scaling factor
when accurate escape modelling is required.

Comparisons with previously used surrogate systems show that reduced-mass scaling based on O--CO$_2$ and
C--CO$_2$ collisions substantially overestimates H--CO$_2$ cross sections, by factors of up to $\sim$45 and
$\sim$30, respectively.
Empirical fits reproduce the general high-energy behaviour but underestimate the low-energy regime by as
much as $\sim$45\%.
Together, these findings underscore the limitations of mass-scaling and fitting approaches widely employed
in escape studies, and highlight the importance of system-specific quantum calculations for light-atom
collisions with polyatomic targets.

\subsection{Implications for atmospheric escape}

The cross sections presented here provide the first quantum-mechanical replacement for the 
simplified collisional parameters that have been used since the foundational escape models of 
\citet{hunten1973escape} and \citet{schunk1980ionospheres}. Our results show that these 
classical approximations overestimate H--CO$_2$ total cross sections by factors of 
30--45 and rate coefficients by approximately 16$\times$. This revision has direct implications 
for modelling early Earth hydrogen escape, where sustained loss is required to explain noble-gas 
isotopic constraints \citep{ZAHNLE201956}. The warmer thermospheres implied by our reduced 
cooling efficiencies could help reconcile escape models with geochemical evidence, demonstrating 
how microscopic quantum scattering physics can reshape macroscopic planetary evolution scenarios.

For thermal (Jeans) escape, the relevant quantity is the momentum-transfer cross section $\sigma_{\rm mt}$.
In a simple isothermal exobase formulation, if a mass-scaled $\sigma_{\rm mt}$ exceeds the true H--CO$_2$
value by a factor $f$, the exobase altitude shifts by
$\Delta z_{\rm exo} \simeq H \ln f$.
Because commonly used mass-scaled values can overestimate the true H--CO$_2$ cross section by up to an
order of magnitude at sub-eV energies, exobase shifts of order $\sim$10--20~km are plausible in such
simplified models.
Since the Jeans flux depends exponentially on the escape parameter, even modest changes in exobase altitude
and temperature can lead to order-unity changes in the thermal escape rate of hydrogen.
Similar considerations apply to deuterium, with the added complexity that its thermalisation efficiency
differs from that of H in a non-uniform, energy-dependent manner.

For non-thermal escape, the connection to the cross sections is more direct.
Photochemically produced hot H and D atoms undergo only a small number of collisions while traversing the
upper thermosphere, so differences of a factor of a few between true and mass-scaled total or
momentum-transfer cross sections translate directly into comparable changes in the number of collisions
experienced by a typical suprathermal atom.
Escape probabilities in Monte Carlo and test-particle models are therefore expected to be systematically
modified when the present state-resolved cross sections are implemented.
Because the isotopic differences in the H--CO$_2$ and D--CO$_2$ cross sections are energy dependent, the
resulting D/H fractionation in escape is likewise expected to differ from current estimates based on
simplified scaling assumptions.

Beyond Mars, the present cross sections are also relevant for CO$_2$ cooling and hydrogen escape in
hydrogen-rich upper atmospheres. Related collisional regimes are also encountered in cometary comae, where CO$_2$ is a major volatile and photodissociation produces suprathermal H atoms in a low-collision environment
\citep{bockelee2004composition,combi2004gas}, and in the present-day terrestrial thermosphere, where
H--CO$_2$ energy transfer contributes to CO$_2$ cooling despite CO$_2$ being a minor constituent
\citep{roble1989co2,mlynczak2010energy}.

In early Earth and early Venus thermospheres, collisional energy transfer between H and CO$_2$ regulates
the efficiency of 15~$\mu$m CO$_2$ cooling, which in turn governs thermospheric temperature structure and
hydrogen loss.
Such processes are central to reconciling noble-gas constraints on early Earth atmospheric evolution, where sustained hydrogen escape is required to explain xenon isotopic fractionation and long-term volatile loss \citep{ZAHNLE201956}, with implications for CO$_2$ cooling efficiency in hydrogen-rich atmospheres \citep{Harman2018}.
The cross sections presented here also provide updated microscopic inputs for revisiting 
scenarios of hydrodynamic escape and noble gas fractionation \citep{ZAHNLE1986462,ZAHNLE1990502}, 
where accurate H--CO$_2$ collision rates are essential for determining diffusion coefficients 
and mass fractionation patterns during the 'flight of the nobility' from early planetary 
atmospheres.

Our calculated momentum-transfer rate coefficients (Table~\ref{tab:rate_coeff}) are approximately 16 times smaller than 
would be obtained by mass-scaling from O--CO$_2$ collisions \citep{gacesa20203}, implying that H--CO$_2$ energy transfer 
in hydrogen-rich regimes may have been less efficient than previously estimated.
This would lead to reduced CO$_2$ cooling efficiency, potentially resulting in warmer upper atmospheres 
and enhanced thermal escape rates --- a factor that could help reconcile models of early atmospheric 
evolution with noble-gas isotopic constraints.
A full assessment of these effects requires coupled thermosphere--ionosphere modelling and is left for
future work.

More broadly, recent progress in planetary escape modelling has shifted major uncertainties from
global drivers such as solar EUV flux toward the microscopic physics of collisional energy and momentum
transfer in the upper atmosphere.
The cross sections presented here therefore provide essential input data for a wide class of escape models,
reducing reliance on mass-scaled or surrogate collision systems and enabling more physically grounded
predictions of atmospheric evolution in CO$_2$-rich planetary and exoplanetary environments.

\section*{Acknowledgments}
This work was supported by Khalifa University of Science and Technology 
(project \#8474000740-RIG-2024-045) and by NASA's Solar System Workings program 
(grant \#22-SSW22-0076). We thank R. Lillis, J. Deighan, M. Chaffin, and B. Gregory for insightful discussions, 
and F. Lique for providing computational resources.

\section*{Conflicts of interest}
There are no conflicts to declare.

\section*{Data availability}
The cross sections and rate coefficients presented in this paper are available 
as electronic supplementary material and have been deposited in the Zenodo 
repository at \url{https://doi.org/10.5281/zenodo.17549249} 
\citep{bop_2025_17549249}.

%%%END OF MAIN TEXT%%%

%%%%%%%%%%%%%%%%%%%% REFERENCES %%%%%%%%%%%%%%%%%%
%\newpage
\bibliographystyle{mnras}
\bibliography{biblio_HCO2_abbr}

%%%%%%%%%%%%%%%%% APPENDICES %%%%%%%%%%%%%%%%%%%%%
%\appendix

%If you want to present additional material which would interrupt the flow of the main paper, it can be placed in an Appendix which appears after the list of references.

%%%%%%%%%%%%%%%%%%%%%%%%%%%%%%%%%%%%%%%%%%%%%%%%%%

% Don't change these lines
\bsp	% typesetting comment
\label{lastpage}
\end{document}